\documentstyle{article}

\font\sqi=cmssq8
\def\DR{\rm I\kern-1.45pt\rm R}
\def\DC{\kern2pt {\hbox{\sqi I}}\kern-4.2pt\rm C}
\begin{document}
\hfill{hep-th/9911020}
\vspace{0.6cm}
\begin{center}
{\Large \bf Massless  particles and the geometry of curves.}\\

 \bigskip
{\Large \bf Classical picture.}\\
 \bigskip
 \vspace{4mm}
{\large A. Nersessian}\footnote{e-mail:nerses@thsun1.jinr.ru}
\\
\smallskip
{\it  Laboratory of Theoretical Physics, JINR,
 Dubna, 141980, Russia}\\
and\\
{\it Department of Theoretical Physics, Yerevan State University,\\
A.Manoukian St., 3, Yerevan, 375049, Armenia}
 \end{center}
\begin{abstract}
 We analyze the  possibility of description
 of $D-$dimensional massless particles by
the Lagrangians linear on world-line curvatures $k_i$,
  ${\cal S}=\sum_{i=1}^Nc_i\int k_i d{\tilde s}$.
 We show, that the nontrivial classical solutions of this model
 are given by  space-like curves with  zero $2N-$th curvature for
 $N\leq[(D-2)/2]$. Massless spinning particles correspond to the curves
 with constant $k_{N+a}/k_{N-a}$ ratio.
It is shown  that only the system with  action
${\cal S}=c\int k_N d{\tilde s}$
leads to irreducible representation of Poincar\'e group.
This system has maximally possible number ($N+1$) of gauge degrees
of freedom. Its  classical solutions obey the conditions
 $k_{N+a}=k_{N-a}$, $a=1,\ldots,N-1$,
while first $N$ curvatures $k_i$ remain arbitrary.
 This solution is  specified  by coinciding $N$
weights of the massless representation of little Lorentz group,
 while the remaining weights vanish.
\end{abstract}
\setcounter{equation}0
\section{Introduction}
 The search of Lagrangians, describing spinning particles, both
 massive and massless, has a long story.
 The conventional  approach in this direction consists
 in the extension of the initial space-time $\DR^{D-1.1}$
 by the auxiliary odd/even coordinates equip a system  with
 spinning degrees of freedom.

 There is another, less developed approach,  where the spinning particle
 systems are described by the Lagrangians,
 which are formulated in the initial space-time, but
 depend on higher derivatives.
 The aesthetically attractive point of the last approach is
 that spinning degrees of freedom are encoded in the geometry
 of its trajectories.
The Poincar\'e and reparametrization invariance
require the   actions to be of the form
\begin{equation}
 {\cal S} =\int {\cal L}({ k}_1,....,{ k}_N)d{\tilde s},
\label{gactions}\end{equation}
 where $k_I$  denote the reparametrization invariants
 (extrinsic curvatures)  of  curves ($0<I\leq D-1 $),
 $d{\tilde s}$ denotes (pseudo)arch length:
 \begin{equation}
 d{\tilde s}=\left\{
\begin{array}{cc}
 |d{\bf x}|&
 {\rm for}\;{\rm non-isotropic}\;{\rm curves}\\
|d^2{\bf x}|^{1/2}&
{\rm for}\;{\rm isotropic}\;{\rm curves}
\end{array}\right.
.
\nonumber\end{equation}
Various systems of this sort, depending on the first and the second
curvatures of path  in  $\DR^{3.1}$ and  $\DR^{2.1}$ are known
(see \cite{misha1,misha2,rr,nr} and refs therein).
Nevertheless, the only system, which leads
to irreducible representation of Poincar\'e group,
is the  model in $\DR^{3.1}$, given by the action\cite{misha1}
$${\cal S}=c\int k_1 d{\tilde s},$$
which describes  the massless spinning particle  with the
helicity $c$ (which, upon quantization,
may take arbitrary integer or half integer values).
Surprisingly, this model has  $W_3$ gauge symmetry \cite{rr},
and is specified by the classical trajectories,
which are  {\it space-like plane curves} with arbitrary
first curvature: $k_1=\forall$, $k_2=0$.

All other three- and four dimensional models lead to the
reducible representations of Poincar\'e group.
For example, the analog of  Plyushchay's model  on isotropic curves
describes massive spinning particles with
Majorana-like spectrum \cite{nr},
    \begin{center}
Spin $\times$ Mass $=c^2/4$.
\end{center}
The subject of this work is to analyze  the problem:

{\it Can massless spinning particles in $D>4$ dimensional
space-time be described by the actions (\ref{gactions})?}\\
In other words, do exist the actions (\ref{gactions}),
which generate the  constraints, corresponding to the  massless
irreducible representations of Poincar\'e group \cite{bala,ss}?

For this purpose  we perform the classical investigation of the
 $D$-dimensional massless (due to explicit scale invariance)  model
on non-isotropic curves given by the action
\begin{equation}
{\cal S}=\sum_{i=1}^{N}c_i\int{k}_id{\tilde s},
\quad c_N\neq 0,  \quad [c_i]=[\hbar],
\label{action}\end{equation}
where
\begin{equation}
d{\tilde s}=|d{\bf x}|,\quad
{ k}_I
=\frac{\sqrt{\det{\hat g}_{I+1}\det{\hat g}_{I-1}}}{\det{\hat g}_I },
 \quad \quad {(g_I)}_{ij}
\equiv
\frac{d^i{\bf x}}{(d{\tilde s})^i}\frac{d^j{\bf x}}{(d{\tilde s})^j},\quad
i,j=1,\ldots,I.
\label{ecurve}\end{equation}
We  restrict ourselves by the Lagrangians linear on curvatures,
since they are specified by the maximally possible (for given $N$) set of
primary constraints \cite{tmp}.
Therefore, they are the only candidates to describe
the massless particle systems, corresponding to irreps.

We establish the following interesting properties of the model (\ref{action}):
\begin{itemize}
\item only the systems with
$$
N\leq N_0=[(D-2)/2],
$$
admit nontrivial classical solutions ($N_0$ is
the rank of the little Lorentz group);
\item the classical solutions of the model under consideration
are space-like curves with $k_{2N}$=0;
the solutions, corresponding to irreps, are specified
by constant $k_{N+a}/k_{N-a}$ ratio, $a=1,\ldots, N-1$;
\item the solutions,
 corresponding to reducible representations, always  exist,
if $\sum_{i=1}^{N}|c_i|>|c_N|$.
\end{itemize}
On the other hand, there is  special case of (\ref{action}),
 exceptional  from many points of view, given by the action
\begin{equation}
         {\cal S}=c\int k_Nd{\tilde s},\quad N\leq N_0=[(D-2)/2]
\label{kN}\end{equation}
The remarable properties of the  model (\ref{kN})
are the following:
\begin{itemize}
\item the solution of the system (\ref{kN})  leads to the irreducible
 representation of Poincar\'e group. This solution are specified
by the following weights of the little Lorentz group:
$m_1=m_2=\ldots=m_N=c,\; m_{N+1}=\ldots=m_{N_0}=0$.
\item the model  (\ref{kN}) has $N+1$ gauge degrees of freedom,
corresponding, probably, classical limit of $W_{N+2}$ algebra.
\item the classical solution of this model is   space-like
     curve specified by the relations:
     $k_1,\ldots k_N$ are arbitrary;
    $k_{N+1}=k_{N-1},\ldots, k_{N+N-1}=k_1$, $k_{2N}=0$.
\end{itemize}
The paper is arranged as follows.\\
In {\it Section 2} we give the Hamiltonian formulation of the
system (\ref{action}) and analyze its general properties.  \\
In {\it Section 3} we present the complete set of constraints
for the model (\ref{kN}) and for the  models  given by
(\ref{action}), where  $N=2,3\;\; c_1c_N\neq 0$ and
$N\geq 3,\;\; c_{i-2}=0,\;c_{N-1}c_N\neq 0,\quad i=3,\ldots N,$.

\setcounter{equation}{0}
 \section{Hamiltonian Formulation}
In order to obtain  the Hamiltonian formulation
 of the system with the action (\ref{action})
we have  to replace it by the classically equivalent one,
which depend on the first-order derivatives,
 and then perform the Legendre
transformation.
For this purpose it is more convenient to use
(instead of explicit expressions (\ref{ecurve}))
the recurrent equations for curvatures,
which follows from the Frenet equations
for moving frame $\{{\bf e}_a\}$:
\begin{eqnarray}
&{\dot{\bf x}}=s{\bf e}_1,\quad {\dot{\bf e}}_a=sK_a^{\;b}{\bf e}_b,
\quad{\bf e}_a{\bf e}_b=\eta_{ab},&\\
& K_{ab}^{\;.}+K_{ba}^{\;.}=0,\quad
K_{ab}^{\;.}=
\left\{\begin{array}{ccc}
\pm k_a,&{\rm if} &b=a\pm 1\\
0,&{\rm if}&b\neq a\pm 1
\end{array}\right.,\quad k_a \geq 0.&
\end{eqnarray}
In the Euclidean space
the Frenet equations
read
\begin{equation}
 {\bf{\dot x}}=s{\bf e}_1,\quad
{\bf\dot e}_a=sk_a{\bf e}_{a+1}-sk_{a-1}{\bf e}_{a-1},\quad\quad
{\bf e}_0={\bf e}_{D+1}\equiv 0,\;\;k_0=k_D=0.
\label{ff}\end{equation}
Consecuently, we get
\begin{equation}
{s}=|{\bf{\dot x}}|={\bf{\dot x}}{\bf e}_1,
\quad sk_i={\bf{\dot e}}_i{\bf e}_{i+1}=
\sqrt{{\bf {\dot e}_i}^2-({ s}k_{i-1})^2 }.
\label{re}\end{equation}
It is easy to verify, that for the transition to the
Frenet equations for non-isotropic curves
in the pseudo-Euclidean space, we do have to substitute,
\begin{equation}
({\bf e}_{\underline a},\; sk_{\underline a},\;
sk_{{\underline a}-1}, s)\to
(i{\bf e}_{\underline a},\; isk_{\underline a},\;
isk_{{\underline a}-1},(-i)^{\delta_{1{\underline a}}}s)
\end{equation}
 for some index ${\underline a}$.\\
The choice ${\underline a}=1$
means the transition to time-like curve,
while ${\underline a}=2,\ldots, D$- to space-like ones.
By this reason, through the paper we  use the
Euclidean signature.

Taking into account the   expressions (\ref{ff}),(\ref{re})
 one can replace the initial Lagrangian (\ref{action})
(in arbitrary time parametrization $d{\tilde s}={s}d\tau$,
${ s}=|{\bf\dot x}|$) by the following one
\begin{eqnarray}
{\cal  L}&={ s}\sum_ic_{i-1}k_{i-1}
 + c_N\sqrt{{\dot{\bf e}}^2_{N} -(sk)^2_{N-1}} +
{\bf p}({\bf{\dot x}}-{ s}{\bf e}_1)
+\sum_{i}{\bf p}_{i-1}
({\dot{\bf e}}_{i-1}- { s}k_{i-1}{\bf e}_{i}+
{ s}k_{i-2}{\bf e}_{i-2})
&\nonumber\\
&-{ s}\sum_{i,j} d_{ij}\left({\bf e}_i{\bf e}_j-\delta_{ij}\right)
&\label{lfo}\end{eqnarray}
where  ${ s}, k_{i-1}, d_{ij}, {\bf p}_{i-1}, {\bf e}_i$
are independent variables, $k_0=0,{\bf p}_0={\bf e}_0=0$.

Performing the Legendre transformation for
this Lagrangian (see for details \cite{tmp}),
 one get the
 Hamiltonian system with the   Hamiltonian structure
\begin{equation}
\begin{array}{c}
\omega_N=d{\bf p}\wedge d{\bf x}+
\sum_{i=1}^N d{\bf p}_i\wedge d{\bf e}_i,\\
{\cal H}={ s}\left[{\bf p}{\bf e}_1+
\sum_{i=1}^{N}{ k}_{i-1}({\phi}_{i-1.i}-c_{i-1})
+\frac{{ k}_{N}}{2c_N}({\Phi}_{N.N}-c^2_N)
+ \sum_{i,j=1}^Nd_{ij}({\bf e}_i{\bf e}_j-\delta_{ij})\right],
\end{array}
\label{ss}\end{equation}
and the  primary  constraints
\begin{eqnarray}
&{\bf p}{\bf e}_1 \approx 0,&\label{phi0}  \\
& {\bf e}_i{\bf e}_j-\delta_{ij}\approx 0,\\
&{\bf p}_N{\bf e}_N\approx 0,\;\;{\bf p}_N{\bf e}_{i-2}\approx 0,
&\label{gauge} \\
& {\phi}_{i-1.i}\equiv {\bf p}_{i-1}{\bf e}_i
- {\bf p}_{i}{\bf e}_{i-1}\approx c_{i-1},\quad
{\Phi}_{N.N}\equiv{\bf p}^2_N-\sum_{i=1}^N({\bf p}_N{\bf e}_i)^2\approx
c^2_N.&\label{primarylinear}
\end{eqnarray}
Notice that in this formulation
 ${ s}$ and ${s}k_i$  play the role of  Lagrangian multipliers,
 so that stabilization of primary constraints generates
either secondary ones, or  the explicit relations on
the first $N$ curvatures.

It is convenient to introduce  the new variables,
instead of ${\bf p}_i$,
\begin{equation}
\begin{array}{c}
{{\bf p}^\bot}_i\equiv{\bf p}_i-
\sum_{j=1}^N({\bf p}_i{\bf e}_j){\bf e}_j, \quad
 {\bf p}^\bot_i{\bf p}^\bot_j\equiv {\Phi}_{i.j},\quad
{{\bf p}^\bot}_i{\bf e}_j=0,\\
{\phi}_{i.j}\equiv{\bf p}_i{\bf e}_j- {\bf p}_j{\bf e}_i,\\
\chi_{ij}={\bf p}_i{\bf e}_j,\quad i\geq j.
\end{array}
\label{pbot}\end{equation}
Since the constraints  $u_{ij}$  are conjugated to $\chi_{ij}$
and commute with  ${\bf p}^\bot_i$ and $\phi_{ij}$,
we can impose, without loss of generality,
the gauge conditions $\chi_{ij}\approx 0$ fixing the values
 of $d_{ij}$
\begin{equation}
\chi_{ij}\approx 0\;:\;\;\Rightarrow \quad
2d_{i.j}= \delta_{ij}({k}_ic_i-{ k}_{i-1}c_{i-1}).
\label{d}\end{equation}
In these variables  the equations of motion
(in  proper-time gauge ${ s}=1$) read
\begin{equation}
\left\{\begin{array}{c}
\dot{\bf x}={\bf e}_1,\\
\dot{\bf e}_{i-1}=k_{i-1}{\bf e}_{i}-k_{i-2}{\bf e}_{i-2},\\
\dot{\bf e}_N=k_N{{\bf p}^\bot}_{N}/c_N-k_{N-1}{\bf e}_{N-1},\\
\dot{{\bf p}^\bot}_i=-\delta_{1.i}{\bf p}-
k_{i-1}{{\bf p}^\bot}_{i-1}+ k_{i+1}{{\bf p}^\bot}_{i+1}-
k_N(\Phi_{i.N}{\bf e}_N-\phi_{i.N}{\bf p}^\bot_N)/c_N\\
{\dot{\bf p}}=0 \\
 {\dot\phi}_{i.j}=-k_{i-1}\phi_{i-1.j}+k_{i}\phi_{i+1.j}+
-k_{j-1}\phi_{i.j-1}+k_{j}\phi_{i.j+1}-
\frac{k_N}{c_N}\delta_{N[i}\Phi_{j].N}
\end{array}\right.
\label{em}\end{equation}
The Poincar\'e generators  of the system take the form
\begin{equation}
{\bf P}={\bf p},\quad {\bf M}= {\bf p}\times{\bf x}
+\sum_{i=1}^N{\bf p}_i\times{\bf e}_i= {\bf p}\times{\bf x} +
\sum_{i=1}^N{{\bf p}^\bot}_i\times {\bf e}_i -\frac 12\sum_{i,j=1}^N
{\phi}_{ij}{\bf e}_i\times{\bf e}_j.
\label{generators}\end{equation}

Now let us construct the secondary  constraints.\\
Stabilization of the constraint (\ref{phi0})
generates the  following set of constraints
\begin{equation}
{\bf p}{\bf e}_i\approx 0,\quad
{\bf p}{\bf p}_i\approx 0 ,
\quad  {\bf p}^2\approx 0,\label{massless}\end{equation}
which provide the model by the  mass-shell
 and transversality conditions.
All  the secondary constraints produced
by the  primary ones (\ref{primarylinear}) are the functions of
$\phi_{ij}$ and $\Phi_{ij}$,
because these functions  commute with $u_{ij}$,(\ref{phi0}) and form
closed algebra.
One can  arrange these functions
in the matrices
$$
 {\hat R}^p_{ij}=(\phi^p_{ij},\Phi^p_{ij}),
\quad
 \phi^p_{ij}\equiv\phi_{i.i+p},\;\;
\Phi^p_{i.j}\equiv\Phi_{i.2N+1-i-p}, \quad p=1,\ldots, N,
$$
so that the secondary constraints of $(p+1)-$th stage,
 depend on   ${\hat R}^{p\pm 1}$.
Thus stabilization procedure contains at most $N$ stage.

 However, the choice of secondary constraints
 is not uniquely defined for any Lagrangian if $N>2$.
The primary constraints (\ref{primarylinear})
 generate the following first-stage secondary constraints,
\begin{equation}
\Phi_{N-1.N}\approx 0,\quad{\phi}_{i-2.i}\approx 0,\quad i=3,\ldots, N.
\label{firststage}
\end{equation}
  At the next stage we get the system of linear equations on $k_i$:
\begin{equation}
\left\{
\begin{array}{c}
 k_{i-4}{\phi}_{i-4.i-1}-k_{i-3}c_{i-2}+
k_{i-2}c_{i-3}-
k_{i-1}{\phi}_{i-3.i} \approx 0,\quad i=3,\ldots, N \\
-k_{N-3}c_N{\phi}_{N-3.N}+k_{N-2}c_Nc_{N-1}-
k_{N-1}c_Nc_{N-2} + k_N\Phi_{N.N-2}\approx 0\\
 k_{N-2}{\Phi}_{N-2.N}+
k_{N-1}({\Phi}_{N-1.N-1}-c^2_N)+k_Nc_Nc_{N-1}\approx 0.
\end{array}\right.
\label{secondstage}\end{equation}
So, only the system with the action (\ref{kN})
give rize to uniquely defined
second-stage  secondary constraints.

{\it Which  is the rule for choosing the secondary constraints,
whicho we have to follow in order to obtain the solution,
corresponding to massless spinning particle?}\\
From (\ref{massless}) it is seen ,
that $N\leq [(N-2)/2]$, while
the helicity matrix  is of the form
\begin{equation}
{\bf S=}\sum_{i=1}^N{{\bf p}^\bot}_i\times {\bf e}_i -\frac 12\sum_{i,j=1}^N
{\phi}_{ij}{\bf e}_i\times{\bf e}_j,\quad
{\rm tr}\;{\bf S}^{2i}= {\rm tr}\;
\left(\begin{array}{cc}
 0&1\\
-{\hat\Phi}&{\hat\phi}\end{array}\right)^{2i}.
\label{helis}\end{equation}
Therefore, only the solutions, containing  $N$ stabilization stages,
 correspond to  irreps.
To get a massless spinning particle system,
we have to choose  the constraints,
which lower the rank of the system  of  the equations linear
in  $k_i$,  and are compatible with the  conditions $k_i\neq 0$. \\

From the equations of motion (\ref{em}) on can see
 that space-like vectors $({\bf e}, {\bf p}^\bot_i)$
define first $2N$ (non-normalized) elements  of moving frame,
while ${\bf p}$ defines  its $(2N+1)-$th,
isotropic element.
One can orthonormalize the vectors ${{\bf p}^\bot}_i$,
introducing \cite{gm}
\begin{equation}
{\bf e}_{N+1}\equiv{\bf p}^\bot_N/c_N,
\quad
 {\bf e}_{2N+1-a}=
-{1\over \sqrt{\Delta_{a} \Delta_{a+1}}}
\left | \begin{array}{ccc}
{\Phi}_{N.N} & \cdots  & {\Phi}_{N.a} \\
\vdots    &         & \vdots  \\
{\Phi}_{a+1 .N } & \cdots  & {\Phi}_{a+1.a} \\
{{\bf p}^\bot}_{N} & \cdots &  {{\bf p}^\bot}_{a},\end{array}
\right |,
\end{equation}
where $\Delta_N=1$,
$\Delta_{a}=\det{\Phi}_{\alpha.\beta}$, $\alpha ,\beta=a,\ldots N$,
$a=1,\ldots, N-1$. \\
Comparing the equations of motion  with (\ref{ff}),
we get the following  relations on curvatures
\begin{equation}
k_{2N}=0,\;\;k_{N+a}
={\sqrt{\frac{\Delta_{N-a}{\Delta}_{N-a+2}}{\Delta^2_{N-a+1}}}}k_{N-a}.
\label{curvcurve}
\end{equation}
Summarizing the results, obtained in this Section,
 we conclude

{\bf Proposition.} The systems with  the actions (\ref{action}) admit
 nontrivial classical solutions, if $N\leq[(D-2)/2]$.
These solutions
are space-like curves with zero $2N-$th extrinsic curvature.
The curves  with constant ratio $k_{N+a}/k_{N-a}$ correspond
to the massless particles with fixed helicities (the classical
analogs of irreps).
The only system, whose solutions corresponds to irreps, is defined
by the action (\ref{kN}).
\setcounter{equation}{0}
\section{Examples}
In this Section we consider the explicit examples of the  systems
defined by the action (\ref{action}).

{\bf Example 1:}${\cal L}=c k_N.$\\
We start with the basic example, given by the action (\ref{kN}),
whose solutions correspond to the massless irreducible
representations of Poincar\'e group.

Primary constraints (\ref{primarylinear}) generate the
maximally possible
set ($N^2$)  of  constraints, all of them  are of the first-class,
\begin{equation}
{\phi}_{ij}\approx 0,\quad
{\Phi}_{ij}-c^2\delta_{ij}\approx 0.
\label{isospinmax}
\end{equation}
The lagrangian multipliers $k_i$ remain arbitrary, hence the system
has $N+1$ gauge degrees of freedom. This is in correspondence
with the conjecture of \cite{rr2} that the gauge symmetries of
the action (\ref{kN}) define the classical limit of $W_{N+2}$ algebra.
Taking into account (\ref{curvcurve}), we conclude,
that classical solution of the system under consideration
is given by space-like curve,
specified by the conditions
$$k_1,\ldots,k_N \;\;{\rm are}\;\;{\rm arbitrary},\quad
{ k}_{N+a}={ k}_{N-a},\;\;a=1,\dots, N-1,\quad k_{2N}=0.$$
The  dimension of  phase space is
 $${\cal D}^{phys}=  2(D-1)+ N(2D-3N-5),$$
so,
\begin{equation}
N_0(2D-3N_0-5)=\left\{
\begin{array}{c}
 (D-2)(D-4)/4,\;\;{\rm for}\; {\rm even}\; D, \\
 (D-1)(D-5)/4,\;\;{\rm for}\; {\rm odd}\; D.
\end{array}
\right.
\end{equation}
Let  introduce the complex variables
$$ {\bf z}_i=({\bf p}_i+\imath {c}{\bf e}_i)/{\sqrt{2}}$$
 in which  the Hamiltonian system reads
\begin{equation}
\begin{array}{c}
\omega=d{\bf p}\wedge d{\bf x}+\frac{\imath}{c}
\sum_{i}d{\bf z}_i\wedge d{\bar{\bf z}}_i,\\
\;\;\\
{\cal H}=\frac{s}{2c}\left[
\imath{\sqrt{2}}{\bf p}({\bf{\bar z}}_1-{\bf{z}}_1)
+\imath\sum_{a=1}^{N-1}{ k}_a({\bf z}_{a}{\bf{ \bar z}}_{a +1}
-{\bf z}_{a+1}{\bf{ \bar z}}_{a})+
{ k}_N({\bf z}_{N}{\bf{\bar z}}_{N}-c^2)\right],
\end{array}
\end{equation}
while the constraints take the conventional form
\begin{equation}
\begin{array}{c}
{\bf z}_i{\bf{\bar z}}_j - c^{2} \delta_{ij}\approx 0, \\
{\bf z}_i{\bf z}_j\approx 0,\\
{\bf p}{\bf z}_i\approx 0, \\
{\bf p}^2\approx 0.
\end{array}
\label{conscomp} \end{equation}
The eigenvalues of the helicity matrix ${\bf S}$,
 $rank {\bf S}=N$,  are given by the relations
${\rm tr}\;{\bf S}^{2i}=c^{2i},\;\; i=1,\ldots N$. \\
So, the system is specified by the following
weights $m_I$ of the little Lorentz group:
$$m_1=\ldots=m_N =c,\quad m_{N+1}=\ldots=m_{N_0}=0.$$
For $N=N_0$ this solution  possesses  conformal  symmetry\footnote
{The author thanks  M.Vasiliev for this remark}\cite{siegel}.\\

{\bf Example 2: ${\cal L}=c_1k_1+c_2k_2, c_{1}\neq 0.$}\\
For this system the constraints (\ref{primarylinear})
produce the secondary constraint $\Phi_{1.2}\approx 0$
and the relation on the curvatures
$$
 k_2c_2c_1+k_1({\Phi}_{1.1}-c^2_2)= 0,
$$
The dimension of phase space is ${\cal D}=2(3D-10)$,
while  the value of ${\Phi}_{1.1}$ remains unfixed
(this function defines the constant of motion).\\

Let us denote $\Phi_{1.1}\equiv c^2$.
Then the curvatures  obey the conditions
$$
k_2c_2c_1=k_1(c^2_2-c^2),\quad ck_3=c_2k_1, \quad k_4=0$$
while the helicities are defined by  the expressions
$$ {\rm tr}{\bf S}^{2i}=(c^2_2-c^2_1)^i+(c^2-c^2_1)^i,\quad i=1,2.$$

{\bf Example 3: ${\cal L}=\sum_{i=1}^3 c_ik_i$, $c_{1}c_3\neq 0$.}\\
The primary constraints (\ref{primarylinear})
produce the following secondary  constraints of the first and second stage
$$\Phi_{2.3}\approx 0,\quad\phi_{1.3}\approx 0,
\quad\Phi_{1.2}\approx 0,
$$
and the relations on the lagrangian multipliers
\begin{equation}
{\hat A}{\hat k}\equiv\left(
\begin{array}{ccc}
c_3c_{2}              &-c_3c_{1}                 &\Phi_{1.3}\\
{\Phi}_{1.3}          &{\Phi}_{2.2}-c^2_3     &c_3c_{2}
\end{array}\right)
\left(
\begin{array}{c}
k_{1}\\
k_{2}\\
k_3
\end{array}\right)=0
\label{3}\end{equation}
When  ${\rm rank} {\hat A}=2$ the system has
no other secondary constraints, and the helicities of the system
are unfixed.

However, if $c_1c_2c_3\neq0$, one can lower the rank of ${\hat A}$
choosing
\begin{equation}
{\Phi}_{2.2}=c^2_3\mp c_1c_3,\quad{\Phi}_{1.3}=\pm c_2c_3,
\label{third}\end{equation}
so  the only preserved relation on curvatures is
\begin{equation}
c_2(k_1\pm k_3)-c_1k_2=0.
\end{equation}
Stabilizing the constraints (\ref{third})
we get
\begin{equation}
{\Phi}_{1.2}\approx0,\quad
({\Phi}_{1.1}-c^2_3\pm
c_1c_3)k_1\pm
c_2c_3k_2=0
\end{equation}
Thus, the function $\Phi_{1.1}$, being the constant of motion,
 remains arbitrary, and we have {\it two} relations on curvatures.\\

{\bf Example 4: ${\cal L}=c_{1}k_{N-1}+c_2k_N,\;\;c_1\neq 0,\;N>2$.}\\
For this system we consider only those sets of constraints,
 which define the solutions, corresponding to the irreps.

The constraints (\ref{primarylinear}) produce the first-stage
secondary constraints (\ref{firststage}).
Then we get the following set of
the second-stage  secondary constraints
$$
\phi_{i-3.i}\approx 0,$$
and the relations on the lagrangian multipliers
$$
{\hat A}{\hat k}=\pmatrix{
c_2c_{1}& 0 &\Phi_{N-2.N}\cr
\Phi_{N-2.N}&{\Phi}_{N-1.N-1}-c^2_2 &c_2c_{1}\cr }
\pmatrix{
 k_{N-2}\cr
 k_{N-1}\cr
k_{N}\cr
}=0.
$$
To get the solutions corresponding to irreps,
we have to choose the constraints, which
lower the rank of the matrix ${\hat A}$, namely
$$
\Phi_{N-1.N-1}\approx c^2_2,\;\;\Phi_{N.N-2}\approx -c_2c_{1}.
$$
Hence  the only preserved relation on $k_i$ is $k_{N-2}= k_{N}$.

Continuing  stabilization procedure,
 we get, finally
\begin{equation}
\left\{
\begin{array}{c}
\phi_{\alpha.\beta}\approx c_{1}\epsilon_{N-\alpha.N-\beta},
\;\;\;\phi_{\alpha.b}\approx 0\;\;
\phi_{a.b}\approx 0,\\
\Phi_{\alpha.\beta}\approx c^2_2\delta_{\alpha\beta},\;\;
\Phi_{\alpha.b}\approx -c_2c_{1}\delta_{\alpha.b+2},\\
\Phi_{a.b}\approx (c^2_{2}-c^2_{1})\delta_{ab}
-c_2c_{1}\delta_{a.b\pm 2},\\
k_{i}= k_{i-2},
\end{array}\right.
\label{ex2}
\end{equation}
where $a,b=1,\ldots N-2;$, $\alpha,\beta=N-1,N$.\\

So, we conclude, that the system under consideration has
the solution, corresponding to irrep.
 Besides the reparametrization and scale invariance,
this solution possesses an  extra gauge  freedom.\\

{\large Acknowledgments.}
The author is  grateful to  M. Vasiliev and S. Lyakhovich
for valuable discussions and comments,
and O. Khudaverdian, H. J. W. M\"ueller-Kirsten,
K. Shekhter, C. Sochichiu for the interest in work useful remarks.

The work has been partially supported by grants INTAS-RFBR No.95-0829,
INTAS-96-538 and the Heisenberg-Landau Program grant HL-99-10.

\end{document}